\documentclass{article}
\usepackage{graphicx}
\graphicspath{ {images/} }
\usepackage[export]{adjustbox}
\usepackage{amsmath}
\usepackage{adjustbox}
\usepackage{float}
\usepackage{authblk}
\usepackage{ragged2e}
\usepackage[bottom]{footmisc}
\usepackage[utf8]{inputenc}

\title{On The Usage Of Average Hausdorff Distance For Segmentation Performance Assessment: Hidden Bias When Used For Ranking}
\author[1]{Orhun Utku Aydin}
\author[2]{Abdel Aziz Taha}
\author[1]{Adam Hilbert}
\author[1,3,4]{Ahmed A. Khalil}
\author[3]{Ivana Galinovic}
\author[3]{Jochen B. Fiebach}
\author[1]{Dietmar Frey}
\author[1,6]{Vince Istvan Madai}

\affil[1]{CLAIM - Charité Lab for AI in Medicine, Charité Universit{\"a}tsmedizin Berlin, Germany }
\affil[2]{Research Studio Data Science, Research Studios Austria, Salzburg, Austria}
\affil[3]{Centre for Stroke Research Berlin, Charité Universitätsmedizin Berlin, Germany}
\affil[4]{Department of Neurology, Max Planck Institute for Human Cognitive and Brain Sciences, Leipzig, Germany}
\affil[5]{Mind, Brain, Body Institute, Berlin School of Mind and Brain, Humboldt-Universität Berlin, Berlin, Germany}
\affil[6]{School of Computing and Digital Technology, Faculty of Computing, Engineering and the Built Environment, Birmingham City University, United Kingdom}

\date{September 2020}

\begin{document}
\maketitle

\begin{center}
\noindent Corresponding Author Details \\
Orhun Utku Aydin \\
orhun-utku.aydin@charite.de \\
\end{center}
\newpage
\tableofcontents
\newpage
\begin{abstract}
Average Hausdorff Distance (AVD) is a widely used performance measure to calculate the distance between two point sets. In medical image segmentation, AVD is used to compare ground truth images with segmentation results allowing their ranking. We identified, however, a ranking bias of AVD making it less suitable for segmentation ranking. To mitigate this bias, we present a modified calculation of AVD that we have coined balanced AVD (bAVD). To simulate segmentations for ranking, we manually created non-overlapping segmentation errors common in cerebral vessel segmentation as our use-case. Adding the created errors consecutively and randomly to the ground truth, we created sets of simulated segmentations with increasing number of errors. Each set of simulated segmentations was ranked using AVD and bAVD. We calculated the Kendall-rank-correlation-coefficient between the segmentation ranking and the number of errors in each simulated segmentation. The rankings produced by bAVD had a significantly higher average correlation (0.969) than those of AVD (0.847). In 200 total rankings, bAVD misranked 52 and AVD misranked 179 segmentations. Our proposed evaluation measure, bAVD, alleviates AVDs ranking bias making it more suitable for rankings and quality assessment of segmentations. 
\end{abstract}

\section{Introduction}
The average Hausdorff Distance (AVD) is a widely used performance measure to calculate the distance between two point sets. In medical image segmentation, AVD is used to compare ground truth images with segmentation results and allows the ranking of different segmentation results. AVD has been applied to assess performance of various applications including brain tumor segmentation (1), cerebral vessel segmentation(2,3), temporal bone segmentation (4), segmentation of the extracranial facial nerve (5), tumor volume delineation (6), and for pylorus tracking in ultrasound images (7). 

AVD is especially recommended for segmentation tasks with complex boundaries and small thin segments such as cerebral vessel segmentation(8). In comparison to other performance measures such as the Dice coefficient, AVD has the advantage that it takes voxel localization into consideration. Unlike the Hausdorff Distance which quantifies the largest segmentation error, AVD takes all distances of point pairs between two segmentations into account.  

In this work, we show, however, that a ranking bias in the usage of AVD makes it less suitable for segmentation ranking. We also present a modified calculation of AVD, coined balanced AVD (bAVD), to alleviate the ranking bias.

\section{Methods}
\subsection{The Average Hausdorff Distance}
The Average Hausdorff Distance (AVD) between two finite point sets X and Y is defined in equation 1.

\begin{equation}
d_{AVD}(X,Y)=(\frac{1}{X} \sum_{x\in X} \min_{y\in Y} d(x,y) + \frac{1}{Y} \sum_{y\in Y} \min_{x\in X} d(x,y)) / 2
\end{equation}

The directed AVD from point set X to Y is given by the sum of all minimum distances from all points from point set X to Y divided by the number of points in X. AVD can be calculated as the mean of the directed AVD from X to Y and directed AVD from Y to X. 
 
In the medical image segmentation domain, the point sets X and Y refer to the ground truth and the segmentation respectively. Equation 1 can be written in a more simplified way as follows:

\begin{equation}
AVD = (\frac{GtoS}{G} + \frac{StoG}{S}) / 2
\end{equation}
where GtoS is the directed AVD from Ground truth to Segmentation, StoG is the directed AVD from Segmentation to ground truth, G is the number of voxels in the ground truth, and S is the number of voxels in the segmentation.

\subsection{The balanced Average Hausdorff Distance}
Since each of the segmentations to be ranked is compared with the ground truth, the bias of AVD stems from the division by S which differs from one segmentation to the other depending on the number of voxels in each segmentation. The modified calculation of AVD is shown in equation (3). Here, StoG is divided by G, which is constant for all segmentations. 
This newly proposed performance measure is coined balanced Average Hausdorff Distance (bAVD). 

\begin{equation}
bAVD = (\frac{GtoS}{G} + \frac{StoG}{G}) / 2
\end{equation}

\subsection{Data}
Time-of-flight MR-Angiography (TOF MRA) images of 10 patients from the 1000Plus study were randomly selected. The 1000Plus study was carried out with approval from the local Ethics Committee of Charite University Hospital Berlin (EA4/026/08). Details about the study have been previously published(9). The only inclusion criteria was no occlusion in any vessel segments constituting the Circle of Willis. To create the ground truth of the cerebral arterial vessels, the 3D TOF MRAs were pre-segmented using a U-net-based deep learning framework and manually corrected by OUA and VIM using ITK-Snap (10) as described in Livne at al. (2).

\subsection{Error simulation}
In order to explore the properties of AVD and bAVD more systematically for quality assessment of cerebral vessel segmentations, an error simulation framework was developed. To simulate segmentations for ranking, a set of 55 non-overlapping segmentation errors common in a vessel segmentation task were manually created. These errors included, for example, over- and undersegmentation of various vessel segments, false positively labeled other anatomical structures and omitted parts of the vessel tree. Of these 55 errors, one random error was added to the ground truth, the image was saved, and this process was successively repeated 9 times by adding each time a new random error to the resulting image and saving each image. The end result was a set of 10 simulated segmentation results with an increasing number of errors in a random combination. 20 such sets were created for each of the 10 patients. 

\subsection{Segmentation Ranking}
AVD and the proposed bAVD were evaluated for their capability to rank the above-generated segmentation sets. Each set of simulated segmentations was ranked using the two performance measures, where the best segmentation result got rank 1 and the lowest rank 10. Here, an ideal performance measure should have a perfect correlation between the produced ranking of simulated segmentations with the increasing number of errors as the next simulated segmentation has at all times an additional error compared to the previous segmentation.

\subsection{Statistical Analysis}
Kendall's tau correlation coefficient for ordinal rankings was calculated between the segmentation rankings of AVD and bAVD and the number of errors in each simulated segmentation set. The arithmetic mean of Kendall's tau correlation coefficients of the 20 ranking sets per patient was calculated. For each patient, the two sided Wilcoxon signed-rank test was performed to calculate whether the improvement of Kendall's tau coefficient was statistically significant between AVD and bAVD. We also analyzed the number of rankings with imperfect correlation with error count, where at least one segmentation was misranked resulting in a correlation score other than 1. 
\newpage
\section{Results}
The rankings produced by bAVD showed significantly higher average Kendall's rank correlation coefficients (0.969) than the rankings of AVD (0.847). In the 200 total rankings analyzed, bAVD led to 52 rankings with at least one misranked segmentation while AVD led to 179. 

An example of rankings produced by AVD and bAVD on an example set of segmentations with increasing number of errors can be found in Table 1. For a complete overview of the averaged results of Kendall's tau coefficients of the 10 patients please see Table 2. 

The new measure bAVD was implemented in the EvaluateSegmentation command line tool that is free to download.\footnote{https://github.com/Visceral-Project/EvaluateSegmentation}

\begin{table}[H]
\textbf{\caption{Example ranking of a set of segmentations with increasing number of errors by AVD and bAVD}}
\begin{adjustbox}{width=1\textwidth}
 \begin{tabular}{| l | l | l | l | l | l |} 
 \hline
 \textbf{Segmentations} & \textbf{Count of Errors} & \textbf{AVD values} & \textbf{AVD rank} & \textbf{bAVD values} & \textbf{bAVD rank} \\ [0.5ex] 
 \hline
 ground\_truth & 0 & 0 & 1& 0 &1  \\ 
 \hline
  E1 & 1 & 0.308 & 2& 0.314 & 2 \\
 \hline
  N1\_E1 & 2 & 0.455 & 3 & 0.467 &3 \\
 \hline
 N1\_K3\_E1 & 3 & 9.836 & 5 & 23.487 &4 \\
  \hline
 N1\_H1\_K3\_E1 & 4 & 10.138 & 7& 24.925 &5 \\
  \hline
 N1\_H1\_K3\_E1\_P991 & 5 & 10.111 & 6& 24.928 &6 \\
  \hline
 N1\_H1\_K3\_G1\_E1\_P991 & 6 & 10.213 & 8 & 25.394 &7 \\
  \hline
 N1\_H1\_K3\_G1\_E1\_P991\_M0 & 7 & 10.345 & 9& 25.435 &8 \\
  \hline
N1\_H1\_K3\_G1\_E1\_V2\_P991\_M0 & 8 & 10.638 & 11 & 25.613 &9 \\
  \hline
N1\_H1\_K3\_G1\_E1\_R1\_V2\_P991\_M0 & 9 & 10.628 & 10& 25.690 &10 \\
  \hline
N1\_H1\_K3\_G1\_E1\_R1\_V2\_P991\_C992\_M0 & 10 & 9.768 & 4& 25.843 &11 \\ [1ex] 
 \hline
\end{tabular}
\end{adjustbox}
Values of performance measures AVD and bAVD(in voxels) are shown with the resulting rankings for one example set. The segmentation ranking of bAVD perfectly correlates with the count of errors in the simulated segmentations. The traditional AVD however fails to properly rank the segmentations in line with the number of errors they contain.
\textit{Error abbreviations are given in the segmentations column. Letters stand for error types and numbers 1,2,3 state the intensity levels subtle,moderate and severe respectively. K3: False positive errors in the skull area. C992: increased radius of the carotid artery(false positive voxels). M0: missing M1 segment of the middle cerebral artery(false negative voxels). E1: False positive segmentation of the optical nerve and adjacent fat tissue. N1: False positive segmentation of the middle meningeal artery(false positive voxels). G1: False positive segmentation of the sigmoid sinus. 
V2: False negative segmentation of small vessels. R1: Random voxels added throughout the image (false positives) H1: False positive segmentation of the meninges. P991: Increased radius of the posterior communicating artery(false positives).}
\label{table:data}
\end{table}

\begin{table}[H]
\textbf{\caption{Results of ranking correlation with number of errors for 10 patients}}
\begin{adjustbox}{width=1\textwidth}
\begin{tabular}{ | c | c | c | c | c | c | c | c | c | c | c | c | c | c | c | c |}  
\hline
\multicolumn{3}{|c|}{Patient 1} & \multicolumn{3}{c|}{Patient 2} & \multicolumn{3}{c|}{Patient 3}  & \multicolumn{3}{c|}{Patient 4}  & \multicolumn{3}{c|}{Patient 5} \\ [0.5ex] 
\hline
PM & Tau & Er &  PM & Tau & Er &  PM & Tau & Er &  PM & Tau & Er &  PM & Tau & Er  \\
\hline
bAVD & 0.991 & 3 &  bAVD & 0.978 & 4 &  bAVD & 0.964 & 7 &  bAVD & 0.978 & 7 &  bAVD & 0.951 & 6 \\
\hline
AVD & 0.893 & 17 &  AVD & 0.836 & 17 &  AVD & 0.867 & 18 &  AVD & 0.915 & 18 &  AVD & 0.836 & 17 \\
\hline
\multicolumn{3}{|c|}{p value = 0.00039} & \multicolumn{3}{c|}{p value = 0.00041} & \multicolumn{3}{c|}{p value = 0.00119}  & \multicolumn{3}{c|}{p value = 0.00018}  & \multicolumn{3}{c|}{p value = 0.00096} \\
   \hline
    \multicolumn{3}{|c|}{Patient 6} & \multicolumn{3}{c|}{Patient 7} & \multicolumn{3}{c|}{Patient 8}  & \multicolumn{3}{c|}{Patient 9}  & \multicolumn{3}{c|}{Patient 10} \\
  \hline
   PM & Tau & Er &  PM & Tau & Er &  PM & Tau & Er &  PM & Tau & Er &  PM & Tau & Er  \\
 \hline
   bAVD & 0.978 & 6 &  bAVD & 0.953 & 7 &  bAVD & 0.967 & 5 &  bAVD & 0.962 & 3 &  bAVD & 0.965 & 4 \\
   \hline
   AVD & 0.767 & 18 &  AVD & 0.860 & 18 &  AVD & 0.84 & 18 &  AVD & 0.833 & 19 &  AVD & 0.824 & 19 \\
   \hline
   \multicolumn{3}{|c|}{p value = 0.00019} & \multicolumn{3}{c|}{p value = 0.00128} & \multicolumn{3}{c|}{p value = 0.00064}  & \multicolumn{3}{c|}{p value = 0.00012}  & \multicolumn{3}{|c|}{p value = 0.00013} \\ [1ex] 
   \hline
\end{tabular}
\end{adjustbox}
Averaged Kendall's rank correlation coefficients over the 20 sets per patient. For each patient, the rankings produced by bAVD had statistically significantly higher Kendall rank correlation coefficients compared to rankings of the traditional AVD. Also the bAVD led to less rankings with at least one misranked segmentation compared to AVD as seen in the number of errors (Er) column.
\textit{PM: Performance measure,Tau: Average Kendall rank correlation coefficient, Er: The number of rankings with at least one misranked segmentation within the total number of 20 rankings per patient, p value: P value of the two sided Wilcoxon signed-rank test comparing the results of 20 sets per patient.}
\label{table:data2}
\end{table}
\begin{figure}[H]
    \includegraphics[width=1\textwidth,center]{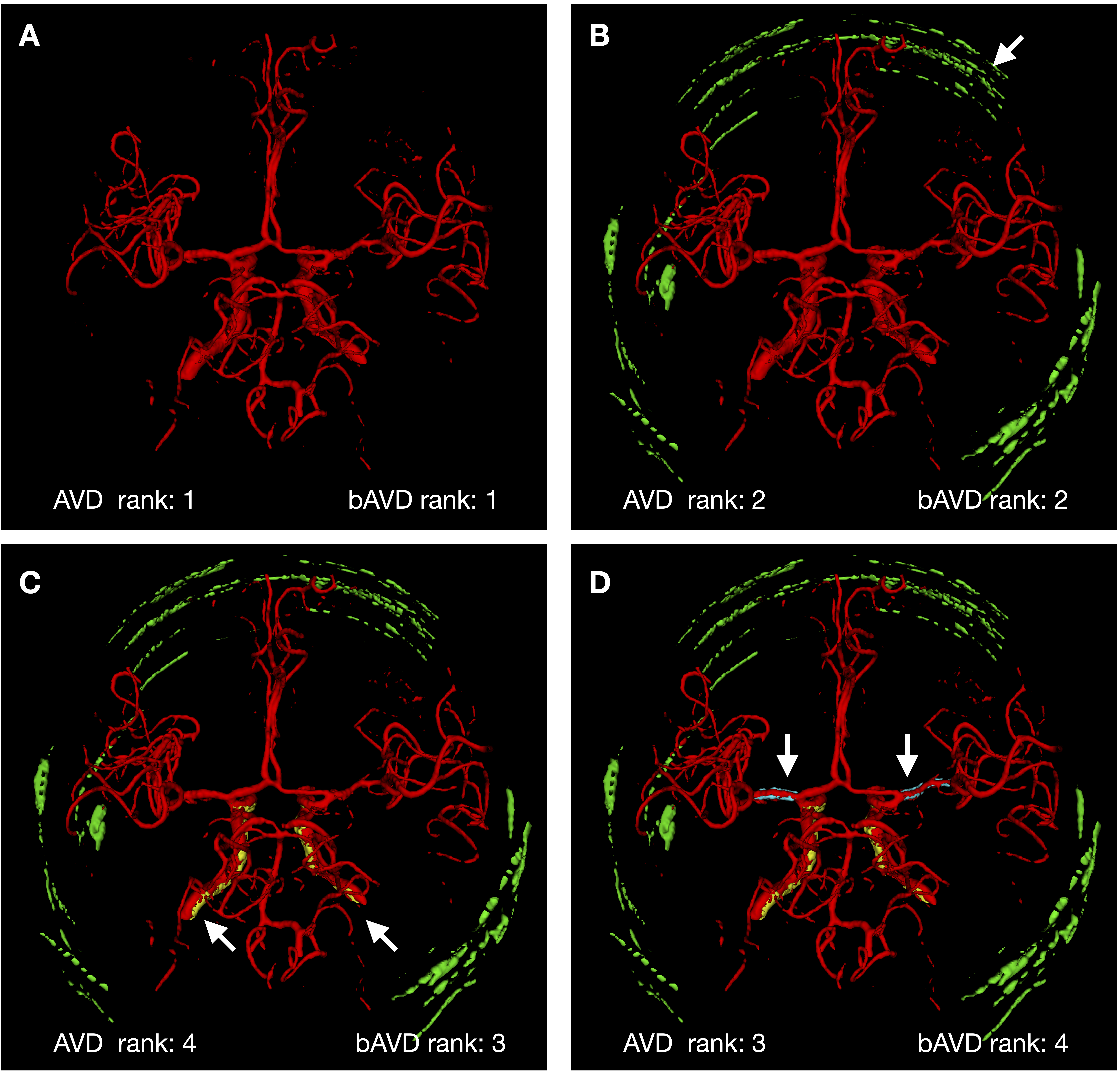}
    \caption{Visual example of the AVD ranking bias. The ground truth image A contains no error, B contains one error, C two errors, and D 3 errors (added errors are indicated with white arrow heads). AVD misranks C and D. In contrast, bAVD correctly ranks all three segmentations according to the number of errors contained in the images. \textit{Green: Error representing false positive voxels in the skull area.Yellow: Error representing voxels removed from the internal carotid arteries bilaterally. Blue: Error representing voxels added to the M1 segment of the middle cerebral artery.} }
\end{figure}

\newpage
\section{Discussion}

Average Hausdorff Distance (AVD) is a recommended and widely-used metric for medical segmentation tasks. In the current paper, we identified a ranking bias of AVD, making it less suitable to compare segmentation results. We also proposed and validated a new metric, bAVD, which strongly alleviates this bias. 

Based on our results, segmentations with lower AVD values do not necessarily correspond to a segmentation of higher quality. Using AVD values to assess segmentation quality may therefore result in an erroneous ranking not reflecting the actual segmentation quality. This ranking bias can be explained by AVD dividing the distance from the ground truth to the segmentation by the number of ground truth voxels while dividing the distance from segmentation to the ground truth by the number of voxels in the segmented volume (Equation 2).This leads to an unwanted bias in certain situations. An additional error added to the segmentation might increase StoG while it simultaneously decreases the AVD indicating an improvement of the segmentation quality. This depends on the distances of the voxels belonging to the error and to the number of voxels contained in the error. Here, although the total distance from segmentation to the ground truth (StoG) increases, the denominator corresponding to the number of voxels in the segmentation volume increases as well. For example, although the simulated segmentation in Figure 1D has an additional error compared to Figure 1C the traditional AVD ranks 1D better than 1C (Figure 1). Therefore, AVD might rank a simulated segmentation containing more errors better than a simulated segmentation with less errors because the denominator changes with the number of voxels in the segmentation. Due to this bias, the traditional AVD should be used with caution for rankings and quality assessment of segmentations. 

This issue was significantly mitigated by the newly proposed bAVG, where the StoG is divided by the constant number of ground truth voxels instead of the variable number of voxels in the segmentation volume. Applying bAVD, the ranking results were strongly improved. However, even with bAVD ranking results were not perfect. There are still a few types of errors that increase StoG and decrease GtoS at the same time. For example, when we simulated false positive single voxels scattered randomly throughout the image volume, we observed increased StoG and decreased GtoS. This resulted in a lower bAVD value, indicating an improved segmentation despite lower quality.

\section{Conclusion}
The novel proposed bAVD performance measure alleviates the identified ranking bias of classic AVD. This makes bAVD more suitable for rankings and quality assessment of segmentations in medical segmentation tasks, and should be used instead of AVD. 
\newpage
\section{Disclosures}
Dr Madai reported receiving personal fees from ai4medicine outside the submitted work. Adam Hilbert reported receiving personal fees from ai4medicine outside the submitted work. Dr Frey reported receiving grants from the European Commission, reported receiving personal fees from and holding an equity interest in ai4medicine outside the submitted work. There is no connection, commercial exploitation, transfer or association between the projects of ai4medicine and the results presented in this work.
\section{References}
\begin{enumerate}
  \item AlBadawy EA, Saha A, Mazurowski MA. Deep learning for segmentation of brain tumors: Impact of cross-institutional training and testing. Med Phys. 2018 Mar;45(3):1150–8.
  \item Livne M, Rieger J, Aydin OU, Taha AA, Akay EM, Kossen T, et al. A U-Net Deep Learning Framework for High Performance Vessel Segmentation in Patients With Cerebrovascular Disease. Front Neurosci [Internet]. 2019 [cited 2019 Jun 2];13. Available from: https://www.frontiersin.org/articles\\ /10.3389/fnins.2019.00097/full
   \item Hilbert A, Madai VI, Akay EM, Aydin OU, Behland J, Sobesky J, et al. BRAVE-NET: Fully Automated Arterial Brain Vessel Segmentation In Patients with Cerebrovascular Disease. Front Artif Intell [Internet]. 2020 [cited 2020 Aug 31];3. Available from: \\https://www.frontiersin.org/articles/10.3389/frai.2020.552258/abstract
   \item Powell KA, Liang T, Hittle B, Stredney D, Kerwin T, Wiet GJ. Atlas-Based Segmentation of Temporal Bone Anatomy. Int J Comput Assist Radiol Surg. 2017 Nov;12(11):1937–44.
   \item Guenette JP, Ben-Shlomo N, Jayender J, Seethamraju RT, Kimbrell V, Tran N-A, et al. MR Imaging of the Extracranial Facial Nerve with the CISS Sequence. AJNR Am J Neuroradiol. 2019;40(11):1954–9.
   \item Peltenburg B, Schakel T, Dankbaar JW, Aristophanous M, Terhaard CHJ, Hoogduin JM, et al. PO-0899: Tumor volume delineation using non-EPI diffusion weighted MRI and FDG-PET in he ad-and-neck patients. Radiother Oncol. 2017 May;123:S496–7.
   \item Chaojie Chen, Yuanyuan Wang, Jinhua Yu, Zhuyu Zhou, Li Shen, Yaqing Chen. Tracking Pylorus in Ultrasonic Image Sequences With Edge-Based Optical Flow. IEEE Trans Med Imaging. 2012 Mar;31(3):843–55.
  \item Taha AA, Hanbury A. Metrics for evaluating 3D medical image segmentation: analysis, selection, and tool. BMC Med Imaging [Internet]. 2015 Dec [cited 2019 Jun 19];15(1). Available from: \\ http://bmcmedimaging.biomedcentral.com/articles/10.1186/s12880-015-\\0068-x
   \item Hotter B, Pittl S, Ebinger M, Oepen G, Jegzentis K, Kudo K, et al. Prospective study on the mismatch concept in acute stroke patients within the first 24 h after symptom onset - 1000Plus study. BMC Neurol. 2009 Dec 8;9:60.
  \item Yushkevich PA, Piven J, Hazlett HC, Smith RG, Ho S, Gee JC, et al. User-guided 3D active contour segmentation of anatomical structures: Significantly improved efficiency and reliability. NeuroImage. 2006 Jul 1;31(3):1116–28.

\end{enumerate}

\end{document}